\documentclass [12pt] {article}

\usepackage{amsfonts}
\usepackage{amssymb}
\usepackage{graphicx} 
\usepackage{hyperref}

\title{Autoresonance in a Dissipative System}

\author{Sergei Glebov \thanks{Ufa State Petroleum Technical University
                              ({\tt sg@anrb.ru})
                             }
        \and
        Oleg Kiselev \thanks{Institute of Mathematics, USC RAS
                             ({\tt ok@ufanet.ru})
                            }
        \and
        Nikolai Tarkhanov \thanks{Institute of Mathematics, Potsdam University
                                  ({\tt tarkhanov@math.uni-potsdam.de})
                                 }
        }

\date{November 12, 2009}

\begin{document}

\maketitle

\begin{abstract}
We study the autoresonant solution of Duffing's equation in the presence of
dissipation.
This solution is proved to be an attracting set.
We evaluate the maximal amplitude of the autoresonant solution and the time
of transition from autoresonant growth of the amplitude to the mode of fast
oscillations.
Analytical results are illustrated by numerical simulations.

\bigskip

\end{abstract}

\section*{Introduction}
\label{s.Int}
\setcounter{equation}{0}

By the autoresonance is meant the growth of the amplitude of oscillations of a
solution to a nonlinear equation under action of an external oscillating force.
This phenomenon looks like phase locking of a nonlinear oscillator through a
periodic driver.
The phase locking was first suggested to accelerate relativistic particles,
see \cite{Veksler,McMillan}.
Nowadays the autoresonance is thought of as universal phenomenon which occurs
in a wide range of oscillating physical systems from astronomical to atomic
ones \cite{FajansFriedland}.

In the general case the frequency of a nonlinear oscillator depends on the
amplitude of oscillations or, what is the same, on their energy.
Hence, in order to change the energy of a nonlinear oscillator the frequency
of the external force should be adopted to that of the oscillator.
If the force is small then the energy of oscillations changes slowly.
In order to remain resonant, the frequency of the external force should adopt
itself also slowly to the frequency of the nonlinear oscillator.
Moreover, the backward phenomenon proves to occur.
Namely, the slow change of the frequency of the driver results in that the
frequency of the nonlinear oscillator actually follows the driver frequency.
For a contemporary survey of the mathematical aspects of autoresonance we refer
the reader to \cite{kalykinReview}.

The autoresonance phenomenon in systems with dissipation was earlier studied
both by means of mathematical models and in physical experiments.
In particular, the existence of autoresonant solution
   for the system of three coupled oscillators with small dissipation was
   established in \cite{YaarivFriedland1993}
and
   for the system with parametric autoresonance in \cite{KhainMeerson2001}.
In the papers
   \cite{FajansGilsonFriedland2001,NaamanAumentadoFriedlandWurteleSiddiqi2008}
the threshold of capture into autoresonance was discussed in the presence of
dissipation.
The resonant phase locking phenomenon in van der Pol Duffing's equation with
external driver of slowly varying frequency was studied in
   \cite{ShagalovEtAll2009}.

In this paper we treat two problems for autoresonance in dissipative systems
for the time being open.
Firstly we prove the existence of an attracting set for solution trajectories
captured into autoresonance.
Such attracting set was observed numerically in a number of papers
   \cite{YaarivFriedland1993,KhainMeerson2001,SKKS,SKKS1}.
The attractor in these systems is a slowly varying steady state solution.
The solutions captured into autoresonance oscillate around such a solution and
lose the energy of oscillations because of dissipation.
Therefore, all captured solutions tend to the steady state solution.
Mathematically this means that the slowly varying steady state solution is
Lyapunov stable.

The second problem we deal with consists in evaluating the bound of the
autoresonant growth of solution in the presence of small dissipation in the
system.
Earlier one observed that the amplitude growth of nonlinear oscillations in
systems with dissipation is bounded, see
   \cite{YaakobiFriedlandHenis,SKKS,SKKS1}.
>From physical viewpoint the boundedness of autoresonant growth can be easily
explained.
Namely, the work of driver is proportional to the length of trajectory in the
phase space.
If the disspation depends linearly on the velocity then its work is proportional
to the area described by the phase trajectory.
Under the growth of energy the area of the phase curve grows faster than its
length.
It follows that even if the dissipation is small, its work exceeds the work of
external force at some moment and the autoresonant growth of solution stops.
Mathematically this looks like the impossibility of extension of the solution
under phase capture.
What happens is the hard loss of stability and passage to fast oscillations.

This work is aimed at finding an asymptotic expansion for the slowly varying
steady state solution to the primary resonance equation and at showing that
it is an attracting set for those solutions which are captured into the
resonance.
Moreover, we derive asymptotics for the maximal amplitude of oscillations
under autoresonance with small dissipation and calculate the period of the
autoresonant mode in the solution.

The paper contains seven sections.
In the next Section \ref{s.sotp} we describe mathematical setting of the
problem.
Section \ref{s.rotp} provides a detailed exposition of the main results.
Section \ref{s.aoag} deals with asymptotics for the autoresonant mode.
In Section \ref{s.soag} we discuss the stability of autoresonant growth.
In Section \ref{s.voboag} we study the solution behaviour in the vicinity of
the break of autoresonant growth.
In Section \ref{s.boag} we will look more closely at the break of autoresonant
growth.
Section \ref{s.fm} is concerned with passage from monotone autoresonant growth
of the amplitude of nonlinear oscillations to fast motion solutions.

\section{Setting of the problem}
\label{s.sotp}
\setcounter{equation}{0}

We study solution of the primary resonance equation
\begin{equation}
\label{pr}
   \imath \mathit{\Psi}'
 + (\tau-|\mathit{\Psi}|^2) \mathit{\Psi}
 + \imath \delta \mathit{\Psi}
 = f,
\end{equation}
where
   $\tau$ is an independent variable,
   $\delta$ a dissipation parameter
and
   $f$ is a parameter related to the amplitude of external force.

The primary resonance equation is of universal character in mathematical
description of autoresonance.
In the case $\delta = 0$ it was first introduced in the paper
\cite{Sinclair}.

The primary resonance equation describes long-term evolution of nonlinear
oscillations under action of a small external force.
As but one example we mention Duffing's equation with dissipation
\begin{equation}
\label{duffing}
   u'' + u + b u' - c u^3 = \varepsilon A \cos (\omega t),
\end{equation}
where
   $c$ and
   $A$ are constants,
   $b$ and $\varepsilon$ small positive parameters.
The frequency of oscillations of the right-hand side of the equation depends
linearly on the time.
More precisely, $\omega = 1 - \alpha t$ which is usually referred to as a
chirped frequency.
The parameter $\alpha$ (called a chirp rate) determines the rate of change of
the frequency.

Duffing's equation (\ref{duffing}) proves to be the simplest and so the most
general equation where one observes the phenomenon of autoresonance break
because of small dissipation.

For studying autoresonance it is convenient to use the method of two scales.
The oscillations of the nonlinear equation are observed in the time scale $t$.
The amplitude of these oscillations depends in turn on the slow time
   $\tau := \varepsilon^{2/3} t$.

The introduction of two time scales enables one to split the evolution of
solution into two parts, fast and slow ones, using the asymptotic substitution
$$
   u
 \sim
   \varepsilon^{1/3} \mathit{\Psi} (\tau) e^{\imath (t-\tau^2)}
 + \mbox{complex conjugate term}
$$
in equation (\ref{duffing}).
The standard averaging procedure over the fast time $t$ in the leading-order term
in $\varepsilon$ leads to equation (\ref{pr}) for the unknown function
$\mathit{\Psi}$, where
   $\delta = \varepsilon^{-2/3} b/4$,
   $f = A/(4 \sqrt{2})$ and
   $c = - 2 \sqrt{2}$.

This primary resonance equation is often written as the system of equations
for the amplitude
   $R (\tau) = |\mathit{\Psi} (\tau)|$
and the phase
   $\varphi (\tau) = \arg (\mathit{\Psi})$.
More precisely,
\begin{equation}
\label{spr}
\begin{array}{rcl}
   R'
 & =
 & -\delta R - f \sin \varphi,
\\
   \varphi'
 & =
 & \displaystyle (\tau-R^2) - \frac{f}{R} \cos \varphi,
\end{array}
\end{equation}
cf. \cite{BM}.

The autoresonance or phase locking for the solution of system (\ref{spr})
means that
   $\varphi' = o (1)$
for $\tau \to \infty$.
This condition along with the second equation of system (\ref{spr}) determine
a low for the amplitude growth which reads
   $R = \sqrt{\tau} + o (1)$.
The first equation of (\ref{spr}) gives a sufficient condition for the instant
at which the phase locking is destroyed, namely
   $\tau_* = f^2 / \delta^2$.
Analysis of the equation and phase locking condition actually yield an estimate
of autoresonance growth in a dissipative system with small dissipation,
   $\delta \ll 1$,
see \cite{YaakobiFriedlandHenis,SKKS,SKKS1}.

In this paper we construct asymptotics for the slowly varying steady state
solution of equation (\ref{spr}) with $\delta \ll 1$.
We prove that this solution is an attracting set for the captured solutions.
Moreover, we show asymptotics of the maximal value of $R$ and evaluate the
instant at which the phase locking is destroyed.

\begin{figure}[h]
\includegraphics [width=7.5cm,keepaspectratio] {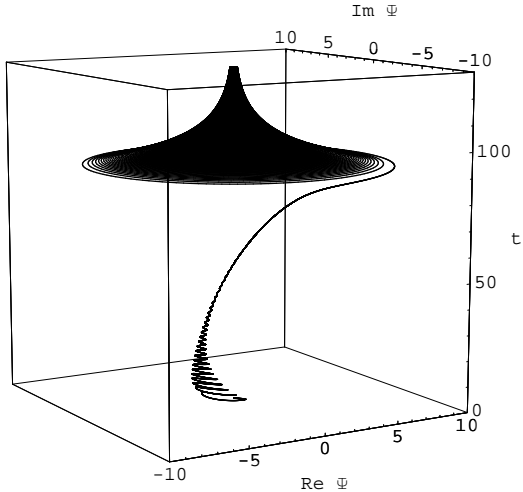}
\caption{%
The solution of (\ref{pr}) with parameters
   $f = 1$,
   $\delta = 0.1$
and the initial condition
   $\mathit{\Psi}\, \restriction_{\tau = 0} = 0$.%
}
\label{fig0}
\end{figure}

In order to better motivate the problem, we demonstrate results of numerical
simulations for equation (\ref{pr}) with $\delta > 0$.
In Figure~\ref{fig0} one can observe three stages of evolution for the solution
of (\ref{pr}).
At the first stage the oscillations are close to some smooth curve.
Then, at the second stage the solution varies slowly.
Finally, at the third stage the solution loses its stability and the amplitude
of fast oscillations tends to zero.

\begin{figure}[h]
\includegraphics [width=7.5cm,keepaspectratio] {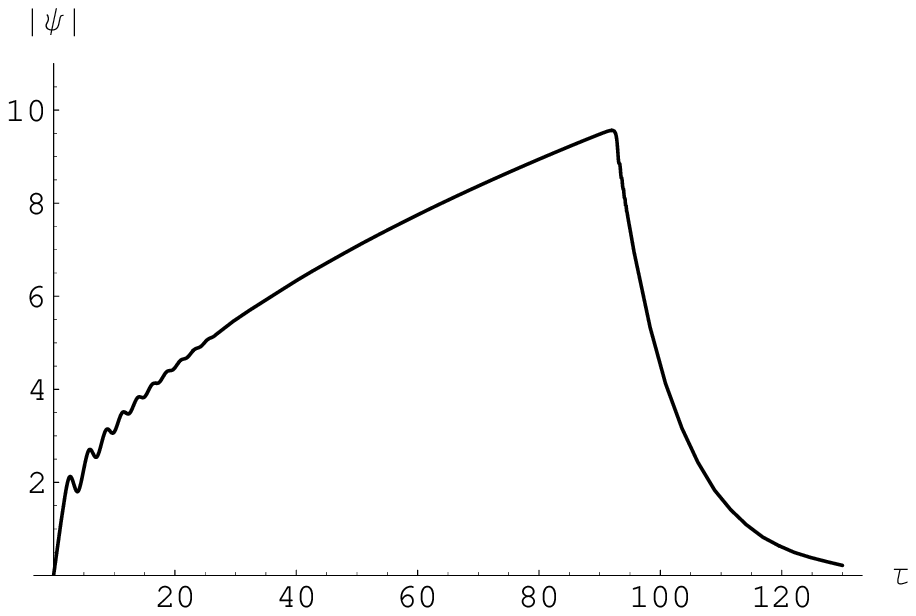}
\caption{%
The modulus of the solution to (\ref{pr}) with parameters
   $f = 1$,
   $\delta = 0.1$
and the initial condition
   $\mathit{\Psi}\, \restriction_{\tau = 0} = 0$.%
}
\label{fig1}
\end{figure}

The autoresonant growth and rapid decay of the amplitude of $\mathit{\Psi}$
are shown in Figure~\ref{fig1}.

\section{Results of the paper}
\label{s.rotp}
\setcounter{equation}{0}

To formulate the results it is convenient to change both the independent and
dependent variables by
\begin{equation}
\label{subst}
\begin{array}{rcl}
   \theta
 & =
 & \tau \delta^2,
\\
   \delta \mathit{\Psi} (\tau)
 & =
 & \psi (\theta,\delta).
\end{array}
\end{equation}
The equation for $\psi$ takes the form
\begin{equation}
\label{perturbedPR}
   \imath \delta^4 \psi'
 + (\theta-|\psi|^2) \psi
 + \imath \delta^3 \psi
 = \delta^3 f.
\end{equation}
Denote
   $R = |\psi|$ and
   $\varphi = \arg (\psi)$
for $\theta > 0$ and
    $0 < \delta \ll 1$.

The existence time for the autoresonant mode in the solution of (\ref{pr}) is
evaluated by
$$
   \theta_*
 = f^2
 - \delta
 + \delta^2 \Big( {f^2 \over 6} z_0 -{1 \over 4 f^2} \Big)
 + O (\sqrt{\delta^5}),
$$
where $z_0$ is the first real pole of the Painlev\'{e}\,-1 transcendental with
zero monodromy data
   $y_1 (z,0,0)$.
Furthermore, the maximal amplitude is estimated by
$$
   R_*
 \sim
   f - {\delta \over 2f} + O (\delta \sqrt{\delta}).
$$

We are now able to give an explicit description of asymptotics for the
autoresonant solution which is an attracting set for the solutions captured
into autoresonance.

If
   $(f^2 - \theta) \delta^{-1} \gg 1$
then $R$ and
     $\varphi$
behave like
\begin{eqnarray*}
   R (\theta,\delta)
 & \sim &
   \sqrt{\theta}
 + \delta^3 {\sqrt{f^2-\theta} \over 2 \theta}
 - \delta^4 {1 \over 2 \theta \sqrt{f^2 - \theta}}
 + \delta^5 {f^2 - 4 \theta \over16 \theta^2 (f^2 - \theta)^{3/2}}
\\
 & + &
   \delta^6
   {(\theta - 3 f^2)(\theta - f^2)^3 + \theta^{3/2} \sqrt{f^2 - \theta}
    \over 8 \theta^{5/2} (f^2 - \theta)^3},
\\
   \varphi (\theta,\delta)
 & \sim &
 - \arctan \Big( {\sqrt{\theta} \over \sqrt{f^2 - \theta}} \Big)
 + \delta {1 \over 2 \sqrt{\theta} \sqrt{f^2 - \theta}}
 + \delta^2 {1 \over 8 \sqrt{\theta} (f^2 - \theta)^{3/2}}
\\
 & + &
   \delta^3
   {(f^2 + 2 \theta) \sqrt{f^2 - \theta} - 24 \sqrt{\theta}(\theta - f^2)^3
    \over 48 (\theta - f^2)^3 \theta^{3/2}}.
\end{eqnarray*}

To write the asymptotics in a neighbourhood of $\theta = f^2$, we change the
variables by
\begin{eqnarray*}
   \eta
 & =
 & (\theta - f^2) \delta^{-1},
\\
   r (\eta,\delta)
 & =
 & (R - f) \delta^{-1},
\\
   a (\eta,\delta)
 & =
 & (\varphi - (3/2) \pi) \delta^{-1/2}.
\end{eqnarray*}
The functions $r (\eta,\delta)$ and
              $a (\eta,\delta)$
have the form
\begin{eqnarray*}
   r
 & \sim &
   {\eta \over 2 f}
 - \delta {\eta^2 \over 8 f^3}
 + \delta^2 {\eta^3 \over 16 f^5}
 - \delta^{5/2} {(2 \eta + 3) \over 4 f^2 \sqrt{-\eta-1}},
\\
   a
 & \sim &
   {\sqrt{-\eta-1} \over f}
 - \delta\, {(2 \eta^2 + 4 \sigma - 1) \over 24 f^3 \sqrt{-\eta-1}},
\end{eqnarray*}
the representations being valid if
   $\delta (-1-\eta)^{-1} \ll 1$.

Close to $\eta = - 1$ it is convenient to represent the asymptotics in the
form
\begin{eqnarray*}
   \tau
 & = &
   {\eta + 1 \over \delta},
\\
   a
 & = &
   \delta^{1/2} u (\tau,\delta),
\\
   r
 & = &
 - {1 \over 2 f}
 + \delta\, {4 f^2 \tau - 1 \over 8 f^3}
 + \delta^2 v (\tau,\delta).
\end{eqnarray*}
In this domain the autoresonant mode of the solution loses its stability.
The leading-order term of $u$ relative to $\delta$ admits the representation
\begin{eqnarray*}
   u (\tau)
 & \sim &
   3^{3/5}\, y_1 (z,0,0),
\\
   \tau
 & = &
   {f^2 \over 6} z - {1 \over 4 f^2},
\end{eqnarray*}
where $y_1 (z,0,0)$ is the Painlev\'{e}\,-1 transcendental,
   see \cite{Kapaev},
i.e., a special solution of the Painlev\'{e}\,-1 equation
   $y_1'' = 6 y_1^2 + z$
with asymptotics
$$
   y_1 (z) = \sqrt{-z \over 6} + O (z^{-1/2}).
$$
The asymptotic formula for $v$ looks like
$$
   v (\tau) \sim {(\tau - u') \over 2 f}
$$
as $\delta \to 0$.

The Painlev\'{e}\,-1 transcendental $y_1 (z,0,0)$ has poles on the real axis.
The approximate solution of (\ref{pr}) by means of $y_1 (z,0,0)$ is valid up
to a small neighbourhood of the first $z_0$ of these poles or,
   what is the same,
up to
   $\tau = \tau_0 := (f^2 / 6) z_0 - 1 / (4 f^2)$.
Near the pole the validity domain is determined by the inequality
$$
   {\sqrt{\delta} \over \tau - \tau_0} \ll 1.
$$

The asymptotics of the solution of (\ref{pr}) in a neighbourhood of the pole
represents by fast non-autoresonant oscillations in the new scale of variable
   $\theta = f^2 - \delta + \delta^2 \tau_0 + \sqrt{\delta^5} \xi$.
It is convenient to write the unknown functions in the form
\begin{eqnarray*}
   R
 & \sim &
   f - {\delta \over 2 f} + \delta \sqrt{\delta}\, p (\xi),
\\
   \varphi
 & \sim &
   {3 \over 2} \pi + s (\xi).
\end{eqnarray*}
The function $s (\xi)$ is a special solution of the equation
$$
   s' = - \sqrt{E + f^2 (s - \sin s)},
$$
such that $s \to 0-$ as
          $\xi \to -\infty$.
The function $p (\xi)$ is determined from the equation
$$
   p = - {s' \over 2 f}.
$$
Note that the function $s (\xi)$ depends on a parameter
   $E = p^2 + (s - \sin s)$
which tends to $0$ as $\xi \to - \infty$.

\section{Asymptotics of autoresonant growth}
\label{s.aoag}
\setcounter{equation}{0}

In this section we construct an asymptotic solution to (\ref{perturbedPR}) in
the domain
   $(f^2 - \theta) \delta^{-1} \gg 1$ and
   $\theta > 0$.
To this end we introduce new unknown functions
   $\rho (\theta,\delta)$ and
   $\alpha (\theta,\delta)$
related to the amplitude $R$ and phase $\varphi$ of the unknown function
$\psi$ by
\begin{eqnarray*}
   R (\theta,\delta)
 & = &
   \sqrt{\theta} + \delta^3 \rho (\theta,\delta),
\\
   \varphi (\theta,\delta)
 & = &
   \alpha (\theta,\delta).
\end{eqnarray*}
On
   substituting these formulas into (\ref{perturbedPR}) and
   separating the real and imaginary parts of the equation
we get
\begin{equation}
\label{eq-rho-alpha}
\begin{array}{rcl}
   \displaystyle
   \delta^4 \rho'
 + \sqrt{\theta}
 + f \sin \alpha
 + \delta {1 \over 2 \sqrt{\theta}}
 + \delta^3 \rho
 = 0,
\\
   (\delta \sqrt{\theta} + \delta^4 \rho) \alpha'
 + 2 \theta \rho
 - f \cos \alpha
 + 3 \delta^3 \sqrt{\theta} \rho^2
 + \delta^6 \rho^3
 = 0.
\end{array}
\end{equation}

Assuming $\delta$ to be small, we look for a solution $\rho$,
                                                      $\alpha$
in the form of asymptotic series
\begin{equation}
\label{series1}
\begin{array}{rcl}
   \rho (\theta,\delta)
 & \sim &
   \displaystyle
   \sum_{k=0}^\infty \delta^k \rho_k (\theta),
\\
   \alpha (\theta,\delta)
 & \sim &
   \displaystyle
   \sum_{k=0}^\infty \delta^k \alpha_k (\theta).
\end{array}
\end{equation}

We first derive equations to determine the coefficients of these asymptotic
series.
For this purpose we substitute (\ref{series1}) into equations
                               (\ref{eq-rho-alpha}).
The trigonometric functions in these equations are expanded as Taylor series
in a neighbourhood of some point $\alpha_0$.
Then we equate the coefficients of the same powers of parameter $\delta$.
As a result we get a recurrent sequence of triangle systems of linear equations
for the unknown coefficients of (\ref{series1}).
In particular, for the leading-order terms of series we obtain
\begin{eqnarray*}
   \sqrt{\theta} + f \sin \alpha_0
 & = &
   0,
\\
   2 \theta \rho_0 - f \cos \alpha_0
 & = &
   0,
\end{eqnarray*}
which gives $\rho_0$ and
            $\alpha_0$.

On equating the coefficients of $\delta$ we arrive at the system
\begin{eqnarray*}
   2 \alpha_1 f \sqrt{\theta} \cos \alpha_0 + 1
 & = &
   0,
\\
   2 \theta \rho_1 + \sqrt{\theta} \alpha_0' + \alpha_1 f \sin \alpha_0
 & = &
   0
\end{eqnarray*}
which readily yields $\alpha_1$ and
                     $\rho_1$
by
\begin{eqnarray*}
   \alpha_1
 & = &
 - {1 \over \sqrt{\theta} \sqrt{f^2 - \theta}},
\\
   \rho_1
 & = &
   {1 \over 2 \sqrt{\theta}}
   \Big( \alpha_1 + {1 \over 2 \sqrt{\theta} \sqrt{f^2 - \theta}} \Big).
\end{eqnarray*}
On equating the coefficients of $\delta^2$ we get the system
\begin{eqnarray*}
 - 2 \alpha_2 \cos \alpha_0 + \sin \alpha_0\, \alpha_1^2
 & = &
   0,
\\
   4 \theta \rho_2
 + 2 \sqrt{\theta} \alpha_1'
 + (\cos \alpha_0\, \alpha_1^2 + 2 \sin \alpha_0\, \alpha_2) f
 & = &
   0
\end{eqnarray*}
implying
\begin{eqnarray*}
   \alpha_2
 & = &
   {\sqrt{\theta} \alpha_1^2 \over 2 \sqrt{f^2 - \theta}},
\\
   \rho_2
 & = &
 - {1 \over 2 \sqrt{\theta}} (\alpha_1' - \alpha_2)
 + {1 \over 4 \theta} \sqrt{f^2 - \theta} \alpha_1^2.
\end{eqnarray*}
On equating the coefficients of $\delta^3$ one still obtains a transparent
system for two unknown functions $\alpha_3$, $\rho_3$
\begin{eqnarray*}
   f \cos \alpha_0\, \alpha_3
 & = &
 - \rho_0
 + {f \over 6}
   \Big( \cos \alpha_0\, \alpha_1^3 + 6 \sin \alpha_0\, \alpha_2 \alpha_1
   \Big),
\\
   2 \sqrt{\theta} \rho_3 - \alpha_3
 & = &
 - \alpha_2'
 - \rho_0^2
 - {f \over \theta} \cos \alpha_0\, \rho_0
 + {1 \over 6} \alpha_1^3
 - {f \over \sqrt{\theta}} \cos \alpha_0\, \alpha_2 \alpha_1
\end{eqnarray*}
whose solution is
\begin{eqnarray*}
   \alpha_3
 & = &
   {1 \over 6} (\alpha_1^3 - 3 \theta)
 - {\sqrt{\theta} \over \sqrt{f^2 - \theta}} \alpha_1 \alpha_2,
\\
   \rho_3
 & = &
 - {1 \over 2 \sqrt{\theta}} (\alpha_2' + \alpha_3)
 - {\sqrt{f^2 - \theta} \over 2 \theta} \alpha_1 \alpha_2
 - {1 \over 12 \sqrt{\theta}} \alpha_1^3
 - \frac{(\theta^2 + 2) (f^2 - \theta)}{8 \sqrt{\theta}},
\end{eqnarray*}
and so on.
\par
Careful analysis of formulas for $\alpha_k$ and
                                 $\rho_k$
obtained in this way actually shows that
\begin{eqnarray*}
   \alpha_k
 & = &
   O ((f^2 - \theta)^{(1-2k)/2}),
\\
   \rho_k
 & = &
   O ((f^2 - \theta)^{(1-2k)/2})
\end{eqnarray*}
as $\theta \to f^2 - 0$.
 From these equalities it follows that the constructed asymptotic expansion is
valid for
   $\delta (f^2 - \theta)^{-1} \ll 1$.

\section{Stability of autoresonant growth}
\label{s.soag}
\setcounter{equation}{0}

We will look for a solution which is a partial sum of the asymptotic series
constructed above, up to remainders
   $\tilde{\rho} (\xi,\theta,\delta)$ and
   $\tilde{\alpha} (\xi,\theta,\delta)$.
Namely, we consider
\begin{equation}
\label{sol1}
   \psi (\theta,\delta)
 = \Big( \sqrt{\theta} + \rho_1 (\theta) \delta^3 + \tilde{\rho} \delta^3
   \Big)
   \exp \Big( \imath (\alpha_0 (\theta)
                    + \alpha_1 (\theta) \delta
                    + \tilde{\alpha} \delta^2)
        \Big)
\end{equation}
where
   $0 < \delta \ll 1$ and
   $\xi = \delta^{-3} \theta$ is the fast variable.

We substitute (\ref{sol1}) into (\ref{perturbedPR}).
The task is now to write down the linear part of the system for
   $\tilde{\rho}$ and
   $\tilde{\alpha}$.
An easy computation yields the system of equations
\begin{equation}
\label{ResTermsSys}
\begin{array}{rcl}
   \tilde{\rho}{}'_{\xi}
 & = &
   \displaystyle
   \Big( \sqrt{f^2 - \theta} \delta - {\delta^2 \over 2 \sqrt{f^2 - \theta}}
   \Big) \tilde{\alpha}
 + f_1 (\theta) + O (\delta^3),
\\
   \tilde{\alpha}{}'_\xi
 & = &
 - 2 \sqrt{\theta} \tilde{\rho} + \delta \tilde{\alpha} + f_2 (\theta)
 + O (\delta^2).
\end{array}
\end{equation}

\begin{figure}[h]
\includegraphics [width=7.5cm,keepaspectratio] {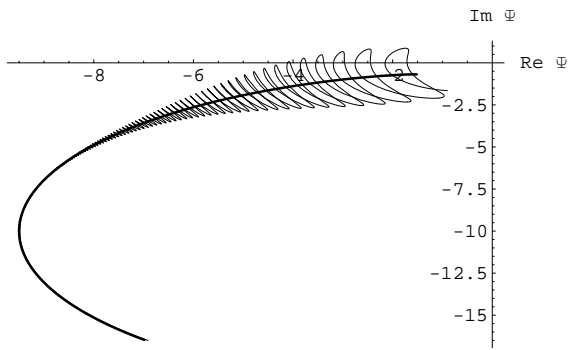}
\caption{%
The graph displays the exponential decay of oscillations with zero initial
data in a neighbourhood of a slowly varying asymptotic solution (bold curve)
with $f = 1$ and
     $\delta = 0.05$.%
}
\label{fig2}
\end{figure}

The right-hand side of the system has coefficients slowly varying in the fast
variable $\xi$.
Solutions of such systems are usually constructed by the WKB method,
   see for instance \cite{Wasow}.
The eigenvalues of the matrix on the right-hand side of (\ref{ResTermsSys})
are
$$
   \lambda_{1,2}
 = \pm \imath\, \sqrt{2} \sqrt[4]{(f^2 - \theta) \theta}\, \delta^{1/2}
 - {\delta \over 2}
 + O (\delta^{3/2}),
$$
and so the real part of eigenvalues is negative.
Hence it follows that the asymptotic solution constructed above is stable in
linear approximation.
Figure \ref{fig2} illustrates this result.

\section{\sloppy Vicinity of the break of autoresonant growth}
\label{s.voboag}
\setcounter{equation}{0}

We change the variables by
   $\theta = f^2 - \delta \eta$.
The new independent variable $\eta$ is stretched with respect to $\theta$.
We will look a solution of the form
\begin{eqnarray*}
   \rho (\theta,\delta)
 & = &
   f + \delta r (\eta,\delta),
\\
   \alpha (\theta,\delta)
 & = &
   {3 \over 2} \pi + \sqrt{\delta}\, a (\eta,\delta).
\end{eqnarray*}
This substitution leads to the system for two unknown functions
   $r (\eta,\delta)$ and
   $a (\eta,\delta)$
\begin{eqnarray*}
   r'
 & = &
 - r + f\, (\cos (\sqrt{\delta} a) - 1)/\delta,
\\
   a'
 & = &
   \delta^{-5/2} (\eta - f r - \delta r^2)
 - \delta^{1/2}\, {f \sin (\sqrt{\delta} a) \over f + \delta r}.
\end{eqnarray*}
This system can be rewritten in a slightly different form
\begin{eqnarray*}
   f a^2
 & = &
 2\, (- r' - r - f\, {\cos (\sqrt{\delta} a) - 1 + \delta a^2/2 \over \delta},
\\
   2 f r
 & = &
   \eta
 - \delta r^2
 - \delta^{5/2}
   \Big( a' - \delta^{-1/2} {f \sin (\sqrt{\delta} a) \over f + \delta r}
   \Big).
\end{eqnarray*}

To find asymptotics we substitute formal series in powers of $\sqrt{\delta}$
for $a (\eta,\delta)$ and
    $r (\eta,\delta)$.
Namely
\begin{eqnarray*}
   r (\eta,\delta)
 & = &
   \sum_{k=0}^\infty r_k (\eta) \delta^{k/2},
\\
   a (\eta,\delta)
 & = &
   \sum_{k=0}^\infty a_k (\eta) \delta^{k/2}.
\end{eqnarray*}
On substituting these series into the system we expand both left-hand side and
                                                            right-hand side
of the equalities as formal series in powers of $\sqrt{\delta}$.
Then we equate the coefficients of the same powers of $\sqrt{\delta}$ in both
series.
As a result we arrive at a recurrent system of equations for determining the
coefficients of formal series.
For $k = 0$ it reads
\begin{eqnarray*}
   2 f r_0
 & = &
   \eta,
\\
   f a_0^2
 & = &
   - 1 - \eta.
\end{eqnarray*}
For $k = 1$ we get $r_1 = 0$ and
                   $a_1 = 0$.
For $k = 2$ the system is
\begin{eqnarray*}
   2 f r_2
 & = &
 - r_0^2,
\\
   f a_0 a_2
 & = &
 -\ r_2' + r_2 + f\, {a_0^4 \over 24},
\end{eqnarray*}
implying
\begin{eqnarray*}
   r_2
 & = &
 - {\eta^2 \over (2 f)^3},
\\
   a_2
 & = &
 - {2 \eta^2 + 4 \eta - 1 \over 24 f^3 \eta + 24 f^3}\, \sqrt{- 1 - \eta},
\end{eqnarray*}
and so on.

The formulas for the coefficients $r_k$ and $a_k$ are cumbersome.
However, using the recurrence relations one can see that the coefficients have
a singularity at $\eta = -1$.
The greater $k$, the higher the singularity.
This is caused
   by differentiating the square root $\sqrt{-1-\eta}$ and
   by increasing the nonlinear dependence on lower order terms of asymptotics
at each step of iteration.
More precisely, we get
\begin{eqnarray*}
   a_k
 & = &
   O ((-1-\eta)^{(1-k)/2}),
\\
   r_k
 & = &
   O ((-1-\eta)^{(4-k)/2})
\end{eqnarray*}
as $\eta \to -1$, provided $k-4 \in \mathbb{N}$.
Hence it follows that the constructed series is asymptotic for
   $\delta\, (-1-\eta)^{-1} \ll 1$.

\section{Break of autoresonant growth}
\label{s.boag}
\setcounter{equation}{0}

In a neighbourhood of the point $\eta = -1$ we change the variables by the
formula
   $\eta = - 1 + \tau \delta$.
The new independent variable $\tau$ is fast with respect to the original
variable $\eta$.
The solution of the primary resonance equation is written in the form
\begin{eqnarray*}
   a
 & = &
   \delta^{1/2} u (\tau,\delta),
\\
   \tau
 & = &
 - {1 \over 2 f}
 + {(4 f^2 \tau - 1) \over 8 f^3} \delta
 + \delta^2 v (\tau,\delta).
\end{eqnarray*}
Substituting these formulas into the system of equations for $a$ and $r$,
we immediately obtain
\begin{eqnarray*}
   v'
 - {1 \over \delta^2} \Big( f \cos (\delta u) - f \Big)
 + {\tau \over 2 f}
 - {1 \over 8 f^3}
 + \delta v
 & = &
   0,
\\
   u'
 - 2 f v
 + {1 \over 8 f^4} (4 f^2 \tau - 1)
 + \frac{f \sin (\delta u)}
        {\displaystyle
         f + \delta \Big( - {1 \over 2f}
                          + {\delta \over 8 f^3} (4 f^2 \tau - 1)
                          + \delta^2 v \Big)}
 & &
\\
 -\
   \delta \Big( {1 \over f} v - {1 \over 64 f^6}(4 f^2 \tau - 1)^2 \Big)
 - \delta^2 {1 \over 4 f^3}\, v\, (4 f^2 \tau - 1) - \delta^3 v^2
 & = &
   0.
\end{eqnarray*}

We will look for a formal solution to this system in the form of power series
in $\delta$
\begin{equation}
\label{series3}
\begin{array}{rcl}
   u (\tau,\delta)
 & = &
   \displaystyle
   \sum_{k=0}^\infty u_k (\tau) \delta^k,
\\
   v (\tau,\delta)
 & = &
   \displaystyle
   \sum_{k=0}^\infty v_k (\tau) \delta^k.
\end{array}
\end{equation}
Substituting these series into the equations and expanding the left-hand sides
as power series in $\delta$, we equate the coefficients of the same powers of
$\delta$.
This leads to a recurrent sequence of differential equations for $u_k$ and
                                                                 $v_k$.
In particular, for $u_0$ and
                   $v_0$
we get
\begin{eqnarray*}
   u_0' + 2 f v_0 - {1 \over 2 f^2} \Big( \tau - {1 \over 4 f^2} \Big)
 & = &
   0,
\\
   v_0'
 + {f \over 2} u_0^2 + {1 \over 2 f^2} \Big( \tau - {1 \over 4 f^2} \Big)
 & = &
   0.
\end{eqnarray*}
For $u_1$ and
    $v_1$
the system looks like
\begin{eqnarray*}
   u_1' + 2 f v_1
 & = &
   {1 \over f} v_0
 - u_0
 - {1 \over 4 f^2} \Big( \tau - {1 \over 4 f^2} \Big)^2,
\\
   v_1' + {f \over 2} u_0 u_1
 & = &
 - v_0.
\end{eqnarray*}
For $u_2$ and
    $v_2$
the system is
\begin{eqnarray*}
   u_2' + 2 f v_2
 & = &
   {1 \over f} v_1
 - {1 \over 4 f^3}\, (1 - 4 f^2 \tau)\, v_0 - u_1 - {1 \over 2 f^2} u_0,
\\
   v_2' + {f \over 2}\, u_0 u_2
 & = &
 - v_1 + {f \over 24}\, u_0^4 - {f \over 2}\, u_1^2,
\end{eqnarray*}
and so on.

The system of equations for the leading-order terms reduces to the Painlev\'{e}\,-1
equation.
To see this, let
\begin{eqnarray*}
   u_0
 & = &
   \sqrt[3]{f^2 \over 6}\, y (z,c_1,c_2),
\\
   v_0
 & = &
   \sqrt[3]{9 \over 2 f^7}\, \frac{d}{d \tau} (y (z,c_1,c_2))
 + {z \over 24 f}
 - {1 \over 8 f^7},
\\
   \tau + {1 \over 4 f^2}
 & = &
   {f^2 \over 6}\, z.
\end{eqnarray*}
Then the differentiation of the equation for $u_0$ leads,
   by the second equation,
to the Painlev\'{e}\,-1 equations in the standard form
   $y'' = 6 y^2 + z$.
Here
   $y = y (z,c_1,c_2)$ is the first Painlev\'{e} transcendental,
   $c_1$ and
   $c_2$ are real parameters of the transcendental which are monodromy data,
   cf. \cite{Kapaev},
and
   $z$ is an independent variable.

The solution of the system of equations for the leading-order term is determined
through the first Painlev\'{e} transcendental.
The parameters of the transcendental are defined by making asymptotic
expansions consistent.
To this end we re-expand asymptotic series (\ref{series3}) in terms of the
variable $\tau$ and equate the coefficients of the same powers of $\delta$.
Then we get asymptotics of the coefficients for $\tau \to - \infty$, namely
\begin{eqnarray*}
   u_0
 & = &
 - {1 \over f} \sqrt{-\tau}
 + O \Big( \frac{1}{\sqrt{-\tau}} \Big),
\\
   v_0
 & = &
   {1 \over 4 f^3} \tau - {1 \over 16 f^5}
 + O \Big( \frac{1}{\sqrt{-\tau}} \Big);
\end{eqnarray*}
\begin{eqnarray*}
   u_1
 & = &
   {1 \over 8 f^5}\, \Big( \frac{1}{\sqrt{-\tau}} \Big)
 + O \Big( \frac{1}{\tau} \Big),
\\
   v_1
 & = &
   {1 \over 8 f^3}\, \tau^2 -{3 \over 16 f^3}\, \tau
 + O \Big( \sqrt{-\tau} \Big);
\end{eqnarray*}
and
\begin{eqnarray*}
   u_2
 & = &
   {1 \over 12 f^3}\, \sqrt{-\tau^3}
 + {5 \over 32 f^5}  \sqrt{-\tau}
 + {1 \over 2 f^2}
 + O \Big( \frac{1}{\sqrt{-\tau}} \Big),
\\
   v_2
 & = &
   {3 \over 16 f^5} \tau^2
 - {5 \over 32 f^7} \tau
 + O \Big( \sqrt{-\tau} \Big).
\end{eqnarray*}

The asymptotics of the $k\,$th correction is
\begin{eqnarray*}
\begin{array}{rclcrcl}
   u_{2n}
 & =
 & O (\sqrt{-\tau^{2n+1}}),
 &
 & u_{2n+1}
 & =
 & O (\sqrt{-\tau^{2n-1}}),
\\
   v_{2n}
 & =
 & O (\tau^{n+1}),
 &
 & v_{2n+1}
 & =
 & O (\tau^{n+2})
\end{array}
\end{eqnarray*}
as $\tau \to -\infty$.

The solution $u_0$ and
             $v_0$
with given asymptotics as $\tau \to -\infty$ can be expressed through the
 first Painlev\'{e} transcendental.
The asymptotics of the first Painlev\'{e} transcendental were investigated in
   \cite{HolmesSpence,Kapaev,GromakLukashevich}.
Here it is convenient to make use of the connection of the asymptotics and the
monodromy data
$$
   u_0
 = \sqrt[3]{\frac{f^2}{6}}\, y (z,c_1,c_2) \restriction_{c_1 = 0 \atop
                                                         c_2 = 0},
$$
see \cite{Kapaev}.
Starting with the formula for $u_0$, one obtains easily an expression for
$v_0 (\tau)$ from the first equation of the system for $u_0$ and
                                                       $v_0$.

We now turn to construction of solutions $u_k$ and
                                         $v_k$.
The corresponding homogeneous system is
\begin{equation}
\label{linearizedEq}
\begin{array}{rcl}
   U' + 2 f V
 & =
 & 0,
\\
   \displaystyle
   V' + {f \over2}\, u_0 U
 & =
 & 0.
\end{array}
\end{equation}
Set
$$
   U (\tau) = \sqrt[3]{f^2 \over 6} w (z).
$$
On differentiating the first equation and
   substituting $V'$ into the second equation
we arrive at the linearised Painlev\'{e}\,-1 equation
$
   w'' + 2 u_0 w = 0.
$
The general solution of this equation is known to be a linear combination of
the partial derivatives of the first Painlev\'{e} transcendental in parameters,
i.e.
$$
   w
 = A_1\, \partial_{c_1} y (z,c_1,c_2) + A_2\, \partial_{c_2} y (z,c_1,c_2)
$$
where $A_1$ and
      $A_2$
are arbitrary constants.
The asymptotics of $y (z,c_1,c_2)$ as $z \to -\infty$ implies
\begin{eqnarray*}
   \partial_{c_1} y (z,c_1,c_2) \restriction_{c_1 = 0 \atop
                                              c_2 = 0}
 & = &
   O (z^{-5/8}),
\\
   \partial_{c_2} y (z,c_1,c_2) \restriction_{c_1 = 0 \atop
                                              c_2 = 0}
 & = &
   O (z^{3/8}),
\end{eqnarray*}
see \cite{Kapaev}.

The formulas for corrections $u_k$ and
                             $v_k$
can now be obtained by the method of variation of constants
$$
   \Big( \begin{array}{c}
         u_k
\\
         v_k
         \end{array}
   \Big)
 = \mathit{\Phi} (\tau)
   \Big( \begin{array}{c}
         A_k
\\
         B_k
         \end{array}
   \Big)
 + \mathit{\Phi} (\tau)
   \int_{a}^\tau
   \mathit{\Phi} (\tau')^{-1}
   \Big( \begin{array}{c}
         f_k
\\
         g_k
         \end{array}
   \Big)
   d \tau'.
$$
Here $\mathit{\Phi} (\tau)$ stands for the fundamental matrix of linearised
system (\ref{linearizedEq}) with Wronskian equal to $1$.
By $a$ is meant an arbitrary real constant satisfying
   $a < \tau_0 := (f^2/6) z_0 - 1/(4 f^2)$,
where $z_0$ is the least real pole of the Painlev\'{e} transcendental.
The constants $A_n$ and
              $B_n$
are uniquely determined from the matching condition for asymptotic solutions.

The first Painlev\'{e} transcendental has second order poles on the real axis,
   see \cite{GromakLukashevich}.
In a neighbourhood of $\tau_0$ the constructed asymptotic expansion no longer
holds.
Indeed, we get
\begin{eqnarray*}
   u_0
 & \sim &
   {6 \over f^2 (\tau - \tau_0)^2},
\\
   v_0
 & \sim &
   {6 \over f^3 (\tau - \tau_0)^3}
\end{eqnarray*}
for $\tau$ close to $\tau_0$.

The general solution for the first correction $u_1$ and
                                              $v_1$
can be represented in the form
\begin{eqnarray*}
   u_1
 & = &
   {a_1 \over (\tau - \tau_0)^3}
 + {3 \over f^4 (\tau - \tau_0)^2}
 + O ((\tau - \tau_0)^{-1}),
\\
   v_1
 & = &
   {3 a_1 \over 2 (\tau - \tau_0)^4}
 + {6 \over f^5 (\tau - \tau_0)^3}
 + O ((\tau - \tau_0)^{-2}).
\end{eqnarray*}
Here $a_1$ is one of the solution parameters.
The second independent parameter is contained in the smooth part of asymptotics
remainder.
The parameters of the solution are uniquely determined while one constructs it
by the method of variation of constants.
However, in the expansion in a neighbourhood of the pole $\tau_0$ the parameter
$a_1$ can be included in the pole translation of the leading-order term of order
$\delta$, namely
   $\tau_1 = \tau_0 - \delta f^2 a_1/3$.
As a result the value $\tau_0$ in the expansion of leading-order terms should be
replaced by $\tau_1$ and the expansions for $u_1$ and
                                            $v_1$
become
\begin{eqnarray*}
   u_1
 & = &
   {3 \over f^4 (\tau - \tau_1)^2} + O ((\tau - \tau_1)^{-1}),
\\
   v_1
 & = &
   {6 \over f^5 (\tau - \tau_1)^3} + O ((\tau - \tau_1)^{-2}).
\end{eqnarray*}
Thus, the pole of the leading-order term of asymptotics is defined uniquely up to
$\delta^2$.
More precisely, the pole asymptotics of the perturbed problem is determined by
singling out summands of order
   $(\tau - \tau_1)^{-3}$
in the asymptotics of $u_k$ for $k > 1$.

The order of singularity at the point $\tau = \tau_1$ increases, for the
higher order corrections depend on lower order corrections in a nonlinear way.
For $u_2$ and
    $v_2$
we have
\begin{eqnarray*}
   u_2
 & \sim &
   {- 18 \over 5 f^6 (\tau - \tau_1)^6},
\\
   v_2
 & \sim &
   {- 54 \over 5 f^7 (\tau - \tau_1)^7}.
\end{eqnarray*}
One can show that
\begin{eqnarray*}
\begin{array}{rclcrcl}
   u_{2n-1}
 & =
 & O ((\tau - \tau_1)^{- 2n}),
 &
 & u_{2n}
 & =
 & O ((\tau - \tau_1)^{- 4n - 2}),
\\
   v_{2n-1}
 & =
 & O ((\tau - \tau_1)^{- 2n - 1}),
 &
 & v_{2n}
 & =
 & O ((\tau - \tau_1)^{- 4n - 3})
\end{array}
\end{eqnarray*}
as $\tau \to \tau_1$.

>From the behaviour of $u_k$ and
                      $v_k$
in a neighbourhood of singular point we deduce that the constructed asymptotics
is valid in the domain
$$
   {\sqrt{\delta} \over |\tau - \tau_1|}
 \ll
   1.
$$

\section{Fast motion}
\label{s.fm}
\setcounter{equation}{0}

In a neighbourhood of the singular point $\tau_1$ the behaviour of the solution
changes drastically.
The solution begins to fastly vary.
The new scale of independent variable is now
$$
   \xi ={(\tau - \tau_1) \over \sqrt{\delta}}.
$$
One introduces new dependent variables
   $p (\xi,\delta)$ and
   $s (\xi,\delta)$
by
\begin{equation}
\label{fastSolution}
\begin{array}{rcl}
   \rho
 & =
 & \displaystyle
   f - 2 {\delta \over f} + \delta \sqrt{\delta}\, p (\xi,\delta),
\\
   \alpha
 & =
 & \displaystyle
   {3 \over 2} \pi + s (\xi,\delta).
\end{array}
\end{equation}
The genuine independent variable $\theta$ is related to
    the new independent variable $\xi$
by the formula
$$
   \theta = f^2 - \delta + \delta^2 \tau_1 + \sqrt{\delta^5}\, \xi.
$$

Substituting the expressions for $\rho$,
                                 $\alpha$
                             and $\theta$
into original system (\ref{eq-rho-alpha}) yields a system of equations for
   $p (\xi,\delta$ and
   $s (\xi,\delta)$.
This system is cumbersome and we need not write it here in explicit form.
Using the standard procedure of perturbation theory we look for the leading-order
term of asymptotics in $\delta$ of the form
$$
\begin{array}{rcl}
   p (\xi,\delta) & \sim & p_0 (\xi),
\\
   s (\xi,\delta) & \sim & s_0 (\xi).
\end{array}
$$
For $p_0 (\xi)$ and
    $s_0 (\xi)$
we obtain the system
\begin{eqnarray*}
   p_0' + f (1 - \cos s_0)
 & =
 & 0,
\\
   s'_0 + 2 f p_0
 & =
 & 0.
\end{eqnarray*}
This system admits the conservation low
$$
   E_0 = p_0^2 + (\sin s_0 - s_0).
$$
Making (\ref{fastSolution}) and
       asymptotics in a neighbourhood of the pole
consistent gives a condition for $p_0$ and
                                 $s_0$,
namely both
   $E_0$ and
   $s_0$
vanish as $\eta \to -\infty$.
The solution of the system for $p_0$ and
                               $s_0$
varies fastly and $s_0$ increases infinitely.
The variable $s_0$ stands actually for the argument of the solution in complex
form
$$
   \Big( f - 2 {\delta \over f} + \delta \sqrt{\delta}\, p (\xi,\delta) \Big)
   \exp \Big( \imath {3 \over 2} \pi +s (\xi,\delta) \Big).
$$
Thus, in this mode the phase locking condition fails to hold and the solution
is not autoresonant.

\section*{Conclusion}
\label{s.conclusion}
\setcounter{equation}{0}

In the paper it is shown for Duffing's equation that in the autoresonance
domain there exists an asymptotic solution which is stable in linear
approximation.
We evaluate the break time and the maximal amplitude of autoresonant solution
for the equation with small dissipation.
The break of the autoresonant mode is accompanied by hard loss of stability
and passage to fast oscillations.

\bigskip

{\bf Acknowledgements\,}
The authors are greatly indebted to L. Kalyakin for many stimulating
conversations.
We gratefully acknowledge the many helpful suggestions of L. Friedland,
                                                          M. Shamsutdinov and
                                                          A. Sukhonosov.
The research was supported by
   the RFBR grant 09-01-92436-KE-a,
   the DFG grant TA 289/4-1
and by
   grant 2215.2008.1 for Russian scientific schools.
Analytic calculations were partially performed by means of the program
   GNU Maxima (http://maxima.sourceforge.net)
under the envelope
   GNU TeXmax (http://www.texmacs.org/).
The authors wish to express their thanks to elaborators of these free
softwares.

\end{document}